\documentclass[conference]{IEEEtran}
\usepackage[latin9]{inputenc}
\usepackage{float}
\usepackage{amsmath}
\usepackage{amsthm}
\usepackage{amssymb, amsfonts}
\usepackage{graphicx}
\usepackage{algorithm}
\usepackage[noend]{algpseudocode}

\makeatletter

\providecommand{\tabularnewline}{\\}
\floatstyle{ruled}
\newfloat{algorithm}{tbp}{loa}
\providecommand{\algorithmname}{Algorithm}
\floatname{algorithm}{\protect\algorithmname}

\theoremstyle{plain}
\newtheorem{thm}{\protect\theoremname}
\theoremstyle{definition}
\newtheorem{defn}[thm]{\protect\definitionname}
\theoremstyle{lemma}
\newtheorem{lemma}{Lemma}
\IEEEoverridecommandlockouts
\usepackage{cite}
\usepackage{amsfonts}
\usepackage{textcomp}
\def\BibTeX{{\rm B\kern-.05em{\sc i\kern-.025em b}\kern-.08em
    T\kern-.1667em\lower.7ex\hbox{E}\kern-.125emX}}

\makeatother

\providecommand{\definitionname}{Definition}
\providecommand{\theoremname}{Theorem}

\begin{document}

\title{Large Player games on Wireless Networks}

\author{\IEEEauthorblockN{Prashant Narayanan\IEEEauthorrefmark{1}, Lakshmi Narasimhan Thegarajan\IEEEauthorrefmark{2}}\IEEEauthorblockA{\IEEEauthorrefmark{1}IIT Palakkad\\
Email: 121813002@smail.iitpkd.ac.in}\IEEEauthorblockA{\IEEEauthorrefmark{2}IIT Palakkad\\
Email: lnt@iitpkd.ac.in}}
\maketitle  
\vspace{-2cm}
\begin{abstract}
We consider a scenario where $N$ users send packets to a common access
point. The receiver decodes the message of each user by treating the
other user's signals as noise. Associated with each user is its channel
state and a finite queue which varies with time. Each user allocates
his power and the admission control variable dynamically to maximize
his expected throughput. Each user is unaware of the states, and actions
taken, by the other users. This problem is formulated as a Markov
game for which we show the existence of equilibrium and an algorithm
to compute the equilibrium policies. We then show that when the number
of users exceeds a particular threshold, the throughput of all users
at all the equilibria are the same. Furthermore the equilibrium policies
of the users are invariant as long as the number of users remain above
the latter threshold. We also show that each user can compute these
policies using a sequence of linear programs which does not depend
upon the parameters of the other users. Hence, these policies can
be computed by each user without any information or feedback from
the other users. We then provide numerical results which verify our
theoretical results. 
\end{abstract}

\section{\label{sec:Introduction}Introduction}

There has been a tremendous growth of wireless communication systems
over the last few years. The success of wireless systems is primarily
due to the efficient use of their resources. The users are able to
obtain their quality of service efficiently in a time varying radio
channel by adjusting their own transmission powers. Distributed control
of resources is an interesting area of study since its alternative
involves high system complexity and large infrastructure due the presence
of a central controller. 

Non cooperative game theory serves as a natural tool to design and
analyze wireless systems with distributed control of resources\cite{NonCop}.
In \cite{Lai}, a distributed resource allocation problem using game
theory on the multiple access channel (MAC) is considered. They considered
the problem where each user maximizes their own transmission rate
in a selfish manner, while knowing the channel gains of all other
users. Scutari et al.\cite{Scutari1} \cite{Scutari2} analyzed competitive
maximization of mutual information on the multiple access channel
subject to power constraints. They provide sufficient conditions for
the existence of unique Nash equilibrium. In a similar setup, \cite{Qiao}
showed that for maximizing the effective capacity of each user, there
exists a unique Nash equilibrium. Heikkinen \cite{Heikkinen} analyzed
distributed power control problems via potential games. 

In \cite{Debbah}, consider a MAC model, where each user knows only
their own channel gain and only the statistics of the channel gains
of other users. The problem is formulated as a Bayesian game, for
which they show the existence of a unique Nash equilibrium.  Altman
et al. \cite{Uplink} studied the problem of maximizing throughput
of saturated users (a user always has a packet to transmit) who have
a Markov modeled channel and are subjected to power constraints. They
considered both the centralized scenario where the base station chooses
the transmission power levels for all users as well as the decentralized
scenario where each user chooses its own power level based on the
condition of its radio channel. In \cite{Wiopt1}, the authors showed
the convergence of the iterative algorithm proposed in \cite{Uplink}.
Altman et al. \cite{Globe} later considered the problem of maximizing
the throughput of competetive users in a distributed manner subject to both power
and buffer constraints.

The works considered so far, compute equilibrium policies for games
with fixed number of users. As the number of user increases, the corresponding
equilibrium policies of users change and the complexity of computing
these policies also increases. To overcome these problems, Population
games\cite{sandholm2010population} models the number of users present
in the system as being significantly large, such that each user can
be modeled as a selfish player playing against a continuum of players.
Then one employs techniques such as evolutionary dynamics to compute
the Nash equilibrium policy of a user in this model. Using the framework
of population games, \cite{Chandramani} model a mobile cellular system,
where users adjust their base station associations and dynamically
control their transmitter power to adapt to their time varying radio
channels. 

Another technique to overcome the latter problems was developed in
\cite{MFG_Lasry,MFG_Lions}. Here each user interacts with other players
only through their average behavior called the mean field. Note that
all the users in these models are considered to be interchangeable\cite{sandholm2010population,MFG_Lasry,MFG_Lions}.
An application of mean field modeling in resource allocation was considered
in \cite{Meriaux}, where each user maximizes their own signal to
interference and noise ratio (SINR). The authors showed this problem
as number of users tends to infinity can be modeled as a Mean field
game. 

The authors in \cite{Wiopt1} consider a different form of analysis
for large number of users. In the model they considered, they showed
that once the number of players in their game exceed some fixed threshold,
the Nash equilibrium policies of each user gets fixed and can be precomputed
in linear time. The authors refer to such a policy as an Infinitely
invariant Nash equilibrium (IINE) policy. They further showed each
user requires no information or feedback from other users to compute
these policies.

However the saturated model considered in \cite{Wiopt1} does not take into account the rate at which data packets arrive form the higher layer. Hence using polcies which are optimal in the saturated model may cause arbitarily large queues at the transmitter. The long delays produced by these policies significantly reduce the quality of service of wireless systems.To mitigate the affect of the latter problem, we consider that each users has a finite buffer where the incoming packets are stored before transmission. We also ensure an average queue constraint on each user. The user then must dynamically allocate power and control the number of packets arriving at its buffer to satisfy its average power as well as queue constraints. Thus unlike the saturated scenario, the user actions affect his state transitions. We model this problem as a Constrained Markov decision game with independent state information\cite{Proof}. Besides providing a algorithm to compute Nash equilibria of this game, we also prove the existence of IINE policies.In the saturated scenario, the IINE was computed using a greedy algorithm. Here The IINE is computed by solving a finite number of linear programs(LP), where at each stage, the current LP requires the solutions of the LP's of the previous stages previous stages. The method of proving this result is different and more general as compared to  the saturated scenario.        


{\em Notations:} Let $g_{i}$ denote an element of the set ${\mathcal{G}}_{i}$
of possible values of a certain parameter associated with the $i$th
user. The set ${\mathcal{G}}=\prod_{i=1}^{N}{\mathcal{G}}_{i}$ denotes
the Cartesian product of these sets. We represent $g=(g_{i},\cdots,g_{N})$,
$g\in\mathcal{G}$ as an element of the set $\mathcal{G}$. The set
${\mathcal{G}}_{-i}=\prod_{j=1,j\neq i}^{N}{\mathcal{G}}_{j}$, denotes
the Cartesian product of the sets other than ${\mathcal{G}}_{i}$
. Any element of this set is represented by $g_{-i}=(g_{i},\cdots,g_{i-1},g_{i+1},\cdots,g_{N})$,
$g_{-i}\in\mathcal{G}_{-i}$. $|{\mathcal{G}}|$ denotes the cardinality
of the set ${\mathcal{G}}$.

\section{\label{sec:System-model}System model}

We denote $n$ as the time index of the $n$'th time slot of a discrete
time system model. We represent ${\mathcal{N}}=\{1,2,3,\cdots,N\}$
as the set of users sending messages to a common receiver over a wireless
medium. We assume that the fading channel gain remains constant over
each time slot . We represent by $h_{i}[n]$, the fading channel gain
of the $i$th user in the $n'$th time slot. The channel gain belongs
to a finite, non-negative, ordered set ${\mathcal{H}}_{i}=\{h_{i}^{0},h_{i}^{1},\cdots,h_{i}^{k}\}$,
where $|{\mathcal{H}}_{i}|=r+1$. The finite set of discrete channel
gains is obtained by the quantization of the channel state information\cite{Uplink,Globe,Wiopt1}.
We assume that the fading channel gain process $h_{i}[n]$ is stationary
and ergodic. The $i$th user transmits with power $p_{i}[n]$ in the
$n$th time slot and the value $p_{i}[n]$ belongs to a finite ordered
set ${\mathcal{P}}_{i}=\{p_{i}^{0},p_{i}^{1},\cdots,p_{i}^{l}\}$,
where $|{\mathcal{P}}_{i}|=q+1$. gives The set $\mathcal{P}$ is
quantization of the transmit power levels \cite{Uplink,Globe}. ,
The set ${\mathcal{P}}_{i}$ includes zero, i.e., $p_{i}^{0}=0$,
as the user may not transmit any message in a time slot. At time slot
$n$, user $i$ can transmit up to $q_{i}[n]$ packets from his finite
buffer of size $Q_{i}$, i.e the value $q_{i}$ belongs to the set
${\mathcal{Q}}_{i}=\{q_{i}^{0},q_{i}^{1},\cdots,Q_{i}\}$. Also at
time slot $n$, user $i$ receives $w_{i}[n]$ from the higher layer,
according to given independent and identically distributed (i.i.d)
distribution $F_{i}$. the incoming packets may be accepted or rejected
by the user, which is indicated by the variable $c_{i}[n]\in\{0,1\}$,
where $c_{i}=1$ and $c_{i}=0$ indicate acceptance and rejection
respectively. Each user can accept packets until the buffer is full,
while the remaining packets are dropped. We assume that in a given
time slot, all arrivals from the upper layer occur after transmission.
The queue process $q_{i}[n]$ evolves as, 

\textbf{\small{}
\begin{eqnarray}
q_{i}[n+1]=\min([q_{i}[n]+c_{i}[n]w_{i}[n]-1_{\{p_{i}[n]>0\}}]^{+},Q_{i}),
\end{eqnarray}
}where $1_{\mathcal{E}}$ denotes the indicator function of the event
$\mathcal{E}$ and $x^{+}$indicates max$(x,0)$. The set of states
${\mathcal{X}}_{i}$ of user $i$ is the Cartesian product of the
set of channel states ${\mathcal{H}}_{i}$ and the set of queue states
${\mathcal{Q}}_{i}$, i.e. ${\mathcal{X}}_{i}:={\mathcal{H}}_{i}\times{\mathcal{Q}}_{i}$.
The set of actions of user $i$ ${\mathcal{A}}_{i}$ is the Cartesian
product of the set $\{0,1\}$ and the set of transmit power ${\mathcal{P}}_{i}$
, i.e. ${\mathcal{A}}_{i}:=\{0,1\}\times{\mathcal{P}}_{i}$. Any element
of these sets ${\mathcal{X}}_{i}$ and ${\mathcal{A}}_{i}$ are represented
as $x_{i}:=(h_{i},q_{i})$ and $a_{i}:=(c_{i},p_{i}),\;c_{i}\in\{0,1\}$
respectively. Each user has an average power and average queue constraints
of $\overline{P_{i}}$and $\overline{Q_{i}}$ respectively. We assume
that each user knows their instantaneous channel gain and queue state
but is not aware of channel gains, queue state and transmit power
of other users. The message of each user is decoded by treating the
signals of the other users as noise. We assume that only user $i$
and the receiver has complete information about the number of packets
in his buffer and the arrival process $w_{i}[n]$ of user $i$ is
independent of his fading process $h_{i}[n]$. The reward function
associated with user $i$ when $h_{i}$,$q_{i}$ and $p_{i}$ are
the instantaneous channel gain, queue state and transmit power of
the $i$th user, respectively is given by,

\begin{eqnarray}
t_{i}(x,a)\triangleq\log_{2}\biggl(1+\frac{h_{i}p_{i}\cdot1_{\{q_{i}>0\}}}{N_{0}+\sum_{j=1,j\neq i}^{N}h_{j}p_{j}\cdot1_{\{q_{j}>0\}}}\biggr),
\end{eqnarray}
where $N_{0}$ is the receiver noise variance.

\section{\label{sec:Problem-Formulation}Problem Formulation}

Here we define the queue and power allocation policies for each user.
Furthermore we define the time average rewards and constraints for
each user. We then formulate the latter as a Markov game, and show
the existence of Nash equilibria for the game. Each user utilizes
a stationary policy $z_{i}(a_{i}/x_{i}),$ which represents the conditional
probability of using action $a_{i}\in\mathcal{A}_{i}$ at state $x_{i}\in\mathcal{X}_{i}$.
Corresponding to each stationary policy and initial distribution $\beta_{i}$
of user $i$ over the set of states ${\mathcal{X}}_{i}$ , we obtain
a probability distribution called occupation measure $z_{i}(x_{i},a_{i})$
on the Cartesian set $\mathcal{X}_{i}\times\mathcal{A}_{i}$. %

 It is defined as,

\begin{eqnarray}
z_{i}(\beta_{i},u_{i};x_{i},a_{i}):=\lim_{T\rightarrow\infty}\frac{1}{T}\sum_{n=1}^{T}\lambda_{\beta_{i}}^{z_{i}}(x_{i}[n]=x_{i},a_{i}[n]=a_{i}). &  & .\label{Def_Occ_msr}
\end{eqnarray}
Under the assumption of unichain MDP, the occupation measure $z_{i}(\beta_{i},u_{i};x_{i},a_{i})$
is well defined for a stationary policy $u_{i}$ and is independent
of the initial distribution $\beta_{i}$ (Theorem. $4.1$, \cite{CMDPBOOK})
and is related to the corresponding stationary policy as,

\begin{eqnarray}
z_{i}(a_{i}|x_{i})=\frac{z_{i}(x_{i},a_{i})}{\sum_{a_{i}\in{\mathcal{A}}_{i}}z_{i}(x_{i},a_{i})},\text{\,}a_{i}\in\mathcal{A}_{i},\text{\,\ensuremath{x_{i}}\ensuremath{\in\mathcal{X}_{i}}. }\label{Calc_stat}
\end{eqnarray}

{} It can be verified that the above MDP is unichain. In this work,
we shall consider only the occupation measures, as the stationary
policy can be obtained from it using (\ref{Calc_stat}) and refer
to them interchangeably. Given policies $z_{i}$, the average rate
obtained by user $i$ is, 
\begin{eqnarray}
T_{i}(z)=\sum_{x_{i},a_{i}}R_{i}^{z_{-i}}(x_{i},a_{i})z_{i}(x_{i},a_{i}),\label{eq:Avg_Rwrd}
\end{eqnarray}
where the instantaneous rate $R_{i}^{z_{-i}}(x_{i},a_{i})$ of user
$i$ is defined as,

\begin{eqnarray}
R_{i}(x_{i},a_{i})=\sum_{x_{-i}}\sum_{a_{-i}}\bigl(\prod_{l=1}^{N}z_{j}(x_{j},a_{j})\bigr)t_{i}(x,a).\label{eq:Instant_Rwrd}
\end{eqnarray}

Similarly, we define the average power and average queue length under
policy $z_{i}$ for user $i$ respectively as, 

\begin{equation}
P_{i}(z_{i})=\sum_{x_{i},a_{i}}p_{i}\cdotp z_{i}(x_{i},a_{i}),\ Q_{i}(z_{i})=\sum_{x_{i},a_{i}}q_{i}\cdotp z_{i}(x_{i},a_{i})\label{eq:Power_cost}
\end{equation}

Any policy $z_{i}$ which satisfies the user's queue and transmit
power constraints is called a feasible policy. Hence, we define the
set of feasible policies $\mathcal{Z}_{i}$ as,

\begin{align}
\begin{aligned}\mathcal{Z}_{i} & =\end{aligned}
 & \Biggl\{ z_{i}(x_{i},a_{i}),\,x_{i}\in\mathcal{X}_{i},\,a_{i}\in\mathcal{A}_{i}\Bigg|\sum_{(x_{i},a_{i})}z_{i}(x_{i},a_{i})=1,\nonumber \\
 & \sum_{(x_{i},a_{i})}[1_{y_{i}}(x_{i})-P_{x_{i}a_{i}y_{i}}]z_{i}(x_{i},a_{i})=0,\;\forall\;y_{i}\in\mathcal{X}_{i},\nonumber \\
 & P_{i}(z_{i})\leq\overline{P}_{i},\,Q_{i}(z_{i})\leq\overline{Q}_{i},\label{eq:Polyhedron}\\
 & \;z_{i}(x_{i},a_{i})\geq0,\;\forall\;(x_{i},a_{i})\in\mathcal{X}_{i}\times\mathcal{A}_{i},\Biggr\}\nonumber 
\end{align}

Each user selects a feasible policy to maximize his average rate (\ref{eq:Avg_Rwrd}).
Hence, we model this problem as a Non cooperative Markov game. A feasible
policy $z_{i}^{*}\in\mathcal{Z}_{i}$ of user $i$ is called a best
response policy if,

\begin{eqnarray}
T_{i}(z_{i}^{*},z_{-i})-T_{i}(z_{i},z_{-i})\geq0,\forall z_{i}\in\mathcal{Z}_{i}.\label{eq:Best_Rep_LP}
\end{eqnarray}

We represent the set of all such policies as $\mathcal{B}_{i}(z_{-i})$.
This Markov game is represented as the following tuple,

\begin{align*}
{\scriptstyle \Gamma_{\mathcal{N}}=\Biggr[\{N\},\{\mathcal{X}_{i}\}_{i\in\mathcal{N}},\{\mathcal{A}_{i}\}_{i\in\mathcal{N}},\{t_{i}\}_{i\in\mathcal{N}},\{\overline{P}_{i}\}_{i\in\mathcal{N}},\{\overline{Q}_{i}\}_{i\in\mathcal{N}},\{F_{i}\}_{i\in\mathcal{N}}\Biggr]}.
\end{align*}

We define $\epsilon-$Nash equilibrium for this game as follows,
\begin{defn}[\emph{$\epsilon-$Nash Equilibrium}]
 \label{def:-Nash_eqb_defn}A feasible policy $z^{*}\in\mathcal{Z}$
for all users is called an $\epsilon-$Nash equilibrium ($\epsilon-$NE)
if for each user $i\in\mathcal{N}$ and for any feasible policy $v_{i}\in\mathcal{Z}_{i}$,
we have 
\begin{eqnarray}
T_{i}(z^{*})-T_{i}(v_{i},z_{-i}^{*})\geq-\epsilon.\label{eq:NashEqb_defn}
\end{eqnarray}
A policy is called Nash equilibrium if $\epsilon=0.$

In the next section, we prove the existence of a Nash equilibrium
and provide a iterative best response algorithm to compute it.
\end{defn}

\section{Existence and Computation of Nash equilibria.}

The existence of Nash equilibria for this game has been proved in
\cite{Globe}. We now propose a iterative best response algorithm
to compute an equilibrium policy.

\begin{algorithm}
\caption{Iterative Best Response Algorithm}\label{Best_resp_algo}
\begin{algorithmic}[1]
\State  \textbf{Set} iteration index $k=0$
\State  \textbf{Initialize} $z(0)\in \mathcal{Z}$
\State  \textbf{Set} $\epsilon>0$
\While{$|z(k+1)-z(k)|\geq \epsilon$ and $|T(z(k+1))-T(z(k))|\geq \epsilon$} 
   \For{j=1:N}
   \State \emph{Set} $z_j(k+1)\in \mathcal{B}_i(z_{-i}(k))$
   \EndFor
\State $k \gets k+1$   
\EndWhile
\State \textbf{return} $z(k)$\Comment{The best responses are $z(k)$ at stage $k$}
\end{algorithmic}
\end{algorithm}

\subsection{Potential games}
\begin{defn}[\emph{Potential Games}]
\label{Ptnl_fnc_dfn}A potential function $\hat{T}:\mathcal{Z}\longmapsto\mathbb{R}$
for the Markov game $\Gamma_{\mathcal{N}}$ is a function which for
all users $i\in\mathcal{N}$, any pair of policies $(z_{i},\hat{z}_{i})$
of user $i$ and for any multi-policy $z_{-i}$ of users other than
user $i$ satisfies, 
\begin{eqnarray}
T_{i}(z_{i},z_{-i})-T_{i}(\hat{z}_{i},z_{-i})=\hat{T}(z_{i},z_{-i})-\hat{T}(\hat{z}_{i},z_{-i}).\label{Pot_Defn}
\end{eqnarray}
\end{defn}
The next condition can be used to check whether game $\Gamma_{\mathcal{N}}$
has an potential function. 
\begin{thm}[Potential Function]
\label{Ver_ptnl} Suppose there exist a function $t$ such that for
any users $i\in{\mathcal{N}}$, for all state-action pair ($x_{-i},a_{-i}$)
other users and any pair of state action $(x_{i},a_{i})$ and $(\hat{x}_{i},\hat{a}_{i})$
of user $i$ the function $t_{i}(x,a)$ satisfies, 
\begin{eqnarray}
t_{i}(x_{i},a_{i},x_{-i},a_{-i})-t_{i}(\hat{x}_{i},\hat{a}_{i},x_{-i},a_{-i})=\label{Pot_cndtn}
\end{eqnarray}
\begin{eqnarray*}
t(x_{i},a_{i},x_{-i},a_{-i})-t(\hat{x}_{i},\hat{a}_{i},x_{-i},a_{-i}).
\end{eqnarray*}
Then there exist a potential function for the Markov game $\Gamma$.
Furthermore, the function 
\begin{eqnarray}
\hat{T}(z)=\sum_{x\in{\mathcal{X}}}\sum_{{a}\in{\mathcal{A}}}{\biggl[\prod_{l=1}^{N}{z_{l}(x_{l},a_{l})}\biggr]\hat{t}(x,a)}
\end{eqnarray}
is a potential function for the Markov game.
\end{thm}
The next result shows that the Markov game $\Gamma_{\mathcal{N}}$
has a potential function.
\begin{thm}
\label{Chk_ptnl} The game $\Gamma_{\mathcal{N}}$ has a potential
function.
\end{thm}
\begin{IEEEproof}
Define the function $t$ as, 

\begin{equation}
t(x,a)\triangleq\log_{2}\biggl(1+\frac{\sum_{j=1}^{N}h_{j}p_{j}\cdot1_{\{q_{j}>0\}}}{N_{0}}\biggr).
\end{equation}

The reader can verify the condition (\ref{Pot_cndtn}). The proof
then follows from theorem \ref{Ver_ptnl}. 
\end{IEEEproof}
Using the latter results, we show that the iterative best response
algorithm (Algorithm\ref{Best_resp_algo}) will converge to an $\epsilon-$Nash
equilibrium in finite iterations.
\begin{thm}
\label{thm:Algorithm_convergence_THm}
\begin{enumerate}
\item When the error approximation is $\epsilon=0$ and suppose the best
response policy provided by the best response algorithm at each stage
is a vertex of the polyhedron $\mathcal{Z}_{i}$, then the algorithm
computes a Nash equilibrium in finite number of iterations. 
\item When the error approximation is $\epsilon>0$, the best response algorithm
computes an $\epsilon-$ NE in finite steps. 
\end{enumerate}
\end{thm}
\begin{IEEEproof}
The proof is same as proof of theorem $3$ in \cite{Arxiv2}.
\end{IEEEproof}
In general, though the number of Linear programs(LP) required to compute
the NE policies using the iterative best response algorithm is less,
in each such computation, we need to calculate the objective function
for the LP, which quickly becomes computationally expensive. Indeed,
the order complexity of computing the objective function is $O\left(\left(2*(Q_{i}+1)*(L+1)*(K+1)\right)^{N-1}\right)$.
Hence even for moderate number of users, the Iterative best response
algorithm becomes unfeasible in practical amounts of time. In the
next section, we overcome this problem by introducing the concept
of an Infinitely invariant Nash equilibrium(IINE).

\section{Games with large number of users.}

We observe that as the number of the users tends to infinity, the
equilibrium policies of user become fixed. Indeed, the equilibria
policies eventually belong to a fixed set, which we shall characterize
in this section. To do so, we first give the definition of an Infinitely
invariant Nash equilibrium (IINE)\cite{Wiopt1}. 
\begin{defn}
\label{def:IINE_defn} A policy $z_{i}^{*}$ of the $i'$th user is
referred to as an \emph{Infinitely invariant Nash equilibrium} \emph{(IINE)
policy} if for some natural number $N^{*}$ and every finite subsets
of users $\mathcal{N}\subseteq\mathbb{Z}^{+}$ such that $|\mathcal{N}|\geq N^{*}$,
the policy $z_{i}^{*}$ is a Nash equilibrium policy for the game
$\Gamma$ , for all users $i\in\mathcal{N}$.
\end{defn}
The existence of an IINE, ensures that the equilibria policy of each
user remains same as long as the number of users remains beyond the
threshold $N^{*}$. In the next theorems, we show the existence of
an IINE under assumption of interchangeability of users \cite{MFG_Lions,MFG_Lasry}.
We first proceed by defining sets which contain such policies. We
define iteratively the set of $k'$th sensitive policies as,

\begin{equation}
\mathcal{S}_{i}^{k}=\arg\max\big\{ l_{i}^{k}(z_{i})\ \big|\ z_{i}\in\mathcal{S}_{i,\,}^{k}\big\},\label{eq:Kth_sensitive_set}
\end{equation}

where the set $\mathcal{S}_{i}^{0}=\mathcal{Z}_{i}$ and the linear
function 

\begin{equation}
l_{i}^{k}(z_{i})=\sum_{x_{i},a_{i}}(-1)^{k+1}(h_{i}p_{i})^{k}\cdot z_{i}(x_{i},a_{i}).\label{eq:Kth_sensitive_function}
\end{equation}

We now define the set of infinitely sensitive policies $\mathcal{S}_{i}$
of user as

\begin{equation}
\mathcal{S}_{i}=\cap_{k=1}^{\infty}\mathcal{S}_{i}^{k}.\label{eq:Infty_Sensitive_set}
\end{equation}

From (\ref{eq:Kth_sensitive_set}), we observe that 
\begin{equation}
\mathcal{S}_{i}^{k}\in\mathcal{S}_{i}^{k-1}\label{eq:Sensitive_set_prop}
\end{equation}
 . Hence one can restate (\ref{eq:Infty_Sensitive_set}) as $\mathcal{S}_{i}=\lim_{k\rightarrow\infty}\mathcal{S}_{i}^{k}.$
In the next theorem, we show that the set containing all the IINE
policies of user $i$ is precisely the set $\mathcal{S}_{i}$.
\begin{thm}[Necessary and Sufficiency for existence of IINE ]
\label{thm:NAS_IINE} Suppose the set $\mathbb{Z}^{+}$ of strictly
positive integers can be partitioned into finite sets $\mathcal{N}_{1},\mathcal{N}_{2},\cdots,\mathcal{N}_{k}$
such that for all users $i$ and $j$ of a set $\mathcal{N}_{l},\,1\leq l\leq k$,
we have $\overline{P}_{i}=\overline{P}_{j}$,$\overline{Q}_{i}=\overline{Q}_{j}$
$\mathcal{H}_{i}=\mathcal{H}_{j}$, $\mathcal{P}_{i}=\mathcal{P}_{j}$
,$\mathcal{Q}_{i}=\mathcal{Q}_{j}$ and $F_{i}=F_{j}$. Furthermore,
assume there exist a policy $z_{i}\in\mathcal{Z}_{i}$, such that
$l_{i}^{1}(z_{i})>0,$ then all the IINE policies of user $i$ belong
to the set $\mathcal{S}_{i}$. Conversely, every policy of the set
$\mathcal{S}_{i}$ is an IINE policy.
\end{thm}
The previous theorem shows that we can always show there exist an
IINE policy if we can show the set $\mathcal{S}_{i}$ is non-empty.
We show this in the next theorem and also provide a \emph{finite}
sequence of iterated Linear programs to compute one such policy.
\begin{thm}[Existence of IINE]
\label{thm:Existence_IINE} The set of infinitely sensitive policies
$\mathcal{S}_{i}$ is nonempty. Furthermore, if we define $M$ to
be the distinct number of elements in the set $\{h_{i}p_{i}|h_{i}\in\mathcal{H}_{i}\,,\,p_{i}\in\mathcal{P}_{i}\}$,
then $\mathcal{S}_{i}=\mathcal{S}_{i}^{M}.$ 
\end{thm}
Theorem (\ref{thm:Existence_IINE}) shows that rather than solving
infinite linear programs as given in condition (\ref{eq:Infty_Sensitive_set}),
we need to only solve for M (finite) number of linear programs. The
integer $M$ is the maximum number of distinct values of the and $ $,SNR
random variable $X_{i}$ which takes values the set $\{h_{i}p_{i}|h_{i}\in\mathcal{H}_{i},p_{i}\in\mathcal{P}_{i}\}$.
The objective functions (\ref{eq:Kth_sensitive_function}) are only
functions of the users states and actions and do not depend upon any
parameters of the other users. Hence these policies can be precomputed
by each user without knowing any information from other users. When
the number of users in the system does cross the number $N^{*},$
these policies are indeed a Nash equilibrium policies. Hence there
is absolutely no need to use the iterative best response algorithm.
Indeed, at large number of users, the complexity of computing the
equilibrium policies becomes linear.

Now, consider the scenario, where $N\geq N^{*}$ and another new player
joins the system. The new user employs his IINE policy, while the
old users employ their previous IINE policies. Again by the definition
of IINE(\ref{def:IINE_defn}), these policies constitute an equilibrium
for the resulting game of $N+1$ players. Once again, the use of IINE
policies obtains significant reduction in complexity. The next theorem
shows that all the IINE policies are interchangeable.
\begin{thm}[Interchangeability of IINE policies.]
\label{thm:-Interchangeability_IINE} Let $z_{i}$ and $z_{i}^{*}$
represent two distinct IINE policies for each user $i$. Then, for
each user $i,$ $T_{i}(z)=T_{i}(z^{*}).$Acknowledgment
\end{thm}

The latter theorem indicates that the users can employ any one of
their IINE policies. Indeed, as there may exist multiple IINE policies,
however all are equivalent in the sense, that they all provide the
same reward to each user. 

\section{Numerical Results}

In this section, we validate our theoretical results using simulations.
We denote the largest power index and the largest channel index by
$L$ and $K$ respectively. The set of channel states and power values
for each user $i$ is the same is equal to $\left\{ 0,\frac{1}{K},\frac{2}{K},\cdots,1\right\} $and
$\left\{ 0,1,\cdots,L\right\} $respectively. We consider a Markov
fading model with channel state transition probabilities given by
$P(0/0)=\frac{1}{2}$,$P(1/0)=\frac{1}{2}$,$P(K-1/K)=\frac{1}{2}$,$P(K/K)=\frac{1}{2}$
and $P(k-1/k)=\frac{1}{2}$,$P(k+1/k)=\frac{1}{2}$,$P(k/k)=\frac{1}{2}$
$\left(1\leq k\leq K-1\right)$. The Noise variance for each simulation
is fixed to be 1.

The maximum number of admissible packets in the buffer of every user
is fixed to be $Q$. Hence the set of queue states for each user $i$
is $\left\{ 0,1,\cdots,Q\right\} .$ The arrival distribution of every
user is Poisson with parameter $\lambda$. The Power and Queue constraint
for each user is the same and is denoted as $\hat{P}$ and $\hat{Q}$
respectively. For a fixed set of parameters (Scenarios), we shall
use the Best response algorithm (\ref{Best_resp_algo}) to compute
a Nash equilibrium policy for $N_{max}$ games. where the number of
players in the $N$'th game $\left(1\leq N\leq N_{max}\right),$is
$N$ itself. We fix $N_{max}=4$ for all scenarios. 

\begin{table}[tbh]
\caption{Simulation Parameters}
\centering{}\label{Table1} %
\begin{tabular}{|c|c|c|c|c|c|c|c|c|}
\hline 
$Scenario$ & $K$ & $L$ & $Q$ & $\hat{P}$ & $\hat{Q}$ & $\lambda$ & $M$ & $N^{*}$\tabularnewline
\hline 
\hline 
$1$ & $2$ & $2$ & $1$ & $.50000$ & $.500$ & $.49$ & $4$ & $3$\tabularnewline
\hline 
$2$ & $2$ & $3$ & $1$ & $.95000$ & $.500$ & $.49$ & $6$ & $3$\tabularnewline
\hline 
$3$ & $2$ & $3$ & $2$ & $1.5500$ & $1.00$ & $.90$ & $6$ & $3$\tabularnewline
\hline 
$4$ & $3$ & $3$ & $2$ & $1.2800$ & $.650$ & $.60$ & $7$ & $3$\tabularnewline
\hline 
$5$ & $3$ & $3$ & $3$ & $2.1000$ & $1.60$ & $1.5$ & $7$ & $4$\tabularnewline
\hline 
$6$ & $2$ & $3$ & $2$ & $1.5500$ & $.900$ & $1.0$ & $6$ & $2$\tabularnewline
\hline 
7 & $2$ & $3$ & $2$ & $1.7000$ & $.900$ & $1.0$ & $6$ & $1$\tabularnewline
\hline 
\end{tabular}
\end{table}

\begin{figure}[tbh]
\includegraphics[width=0.5\textwidth,height=7cm]{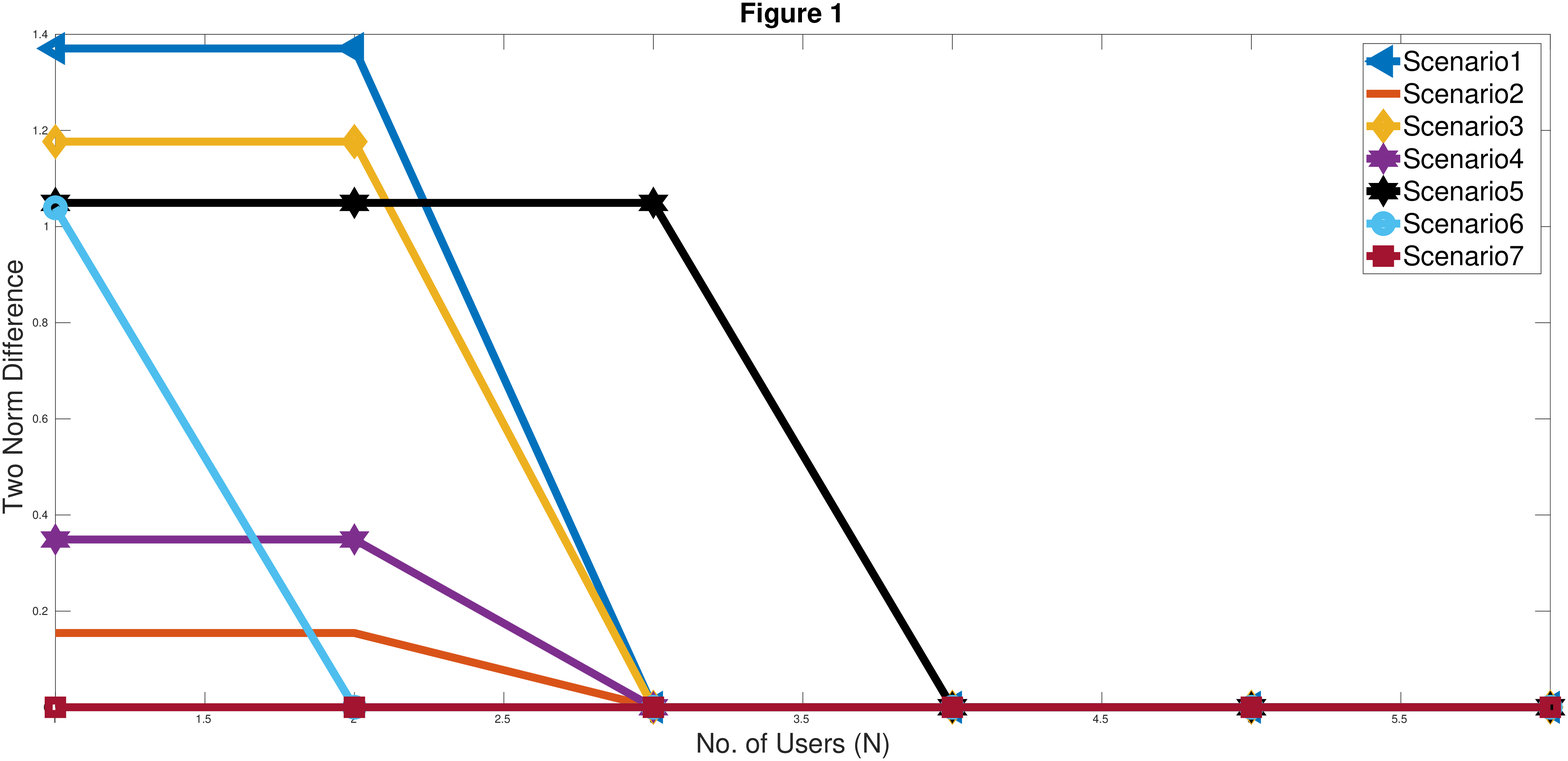}
\raggedright{}\caption{Plot of the $l_{2}$ norm distance between the NE policy and the IINE
policy of the first user against the total number of user as it varies
from $1$ to $N_{max}$.}
\label{Figure1}
\end{figure}

We list the parameters considered in the various scenarios in Table
\ref{Table1}. The last two columns include the the number of linear
programs $\left(M\right)$ required to compute the IINE policy and
the minimum number of users $\left(N^{*}\right)$ at which the IINE
policy becomes an equilibrium policy. In Figure \ref{Figure1}, we
plot the $l_{2}$ norm distance between the NE policies and the IINE
policy of user $1$, as the number of users$(N)$ varies from $1$
to $N_{max}.$ At each value of $N$, the best response algorithm
is used to compute a NE policy ($z_{1}(N)$) of user $1$ for each
scenario. Then in Figure (\ref{Figure1}), we plot $||z_{1}(N)-z_{1}^{*}||_{2}$
for each scenario versus $N$, as $N$ varies from $1$ to $N_{max}$.
$z_{1}^{*}$denotes the invariant policy of user $1$ and is calculated
by solving a sequence of linear programs as given in theorem (\ref{thm:Existence_IINE}).
The $l_{2}$norm between two policies $z_{1}$ and $z_{1}^{*}$is
defined as 

\[
||z_{1}(N)-z_{1}^{*}||_{2}=\sqrt{\sum_{x_{i},a_{i}}\left(z_{1}\left(x_{1},a_{1}\right)-z_{i}^{*}\left(x_{1},a_{1}\right)\right)^{2}.}
\]

We observe from figure \ref{Figure1}, that the NE policy of user
$1$, has become equal to the IINE policy when the number of users
exceed $N^{*}$ as shown in table (\ref{Table1}).

\begin{figure}[tbh]
\includegraphics[width=0.5\textwidth,height=7cm]{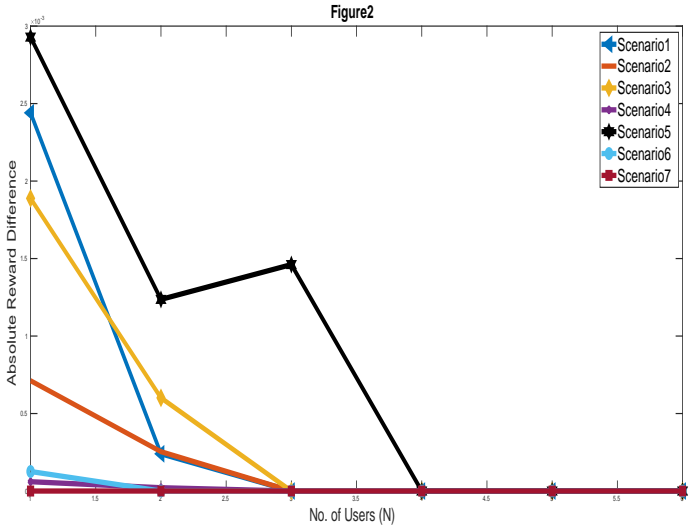}

\caption{Plot of the absolute difference between the reward of user $1$ when
all users use their NE policies and the reward of user $1$ when all
users use their IINE policies against the total number of user as
it varies from $1$ to $N_{max}$.}
\label{Figure2}
\end{figure}

Thus after $N^{*}$, there is no need to use the computationally expensive
Iterative best response algorithm, and rather we can compute the IINE
policy simply. The same result is enforced in figure \ref{Figure2}.
Here we plot the absolute difference between the time average rate
of user $1$, when all the $N$ users use their NE policies and the
time average rate of user $1$, when all the $N$ users use their
IINE policies, against the number of users $N.$ That is, we plot
$|T_{1}(z(N))-T_{1}(z^{*}(N))|$ versus $N$, where $z(N)=\left(z_{1}(N),\cdots z_{N}(N)\right)$
represents the NE policies of all the $N$ users, when there are $N$
players in the game and $z^{*}(N)=\left(z_{1}^{*}(N),\cdots z_{N}^{*}(N)\right)$
represents the IINE polices of the $N$ users. Here also we can see
that after the critical number $N\geq N^{*}$, the equilibrium reward$(T_{1}(z(N)))$
of user $1$ is the same as the reward$(T_{1}(z^{*}(N)))$ when all
the $N$ users employ their IINE policy. Indeed as in these scenarios,
the NE polices have become equal to the IINE policies as shown in
figure \ref{Figure1}, the rewards then also become the same.

From table \ref{Table1}, $N^*$ is $1$ for secnario $7$. This implies that the IINE is 
a NE policy when total number of users exceeds $1$.As there have to be atleast one user, hence for this scenario, the IINE is a optimal solution to the single user problem where the user maximzies their own rate subject to power and queue constraint. Futhermore as shown in Figure \ref{Figure1} and Figure \ref{Figure2} this policy remains a NE policy for each user irrespective of number of users in the system.

\begin{figure}[tbh]
\hspace{-8mm}
\includegraphics[width=0.6\textwidth,height=8cm] {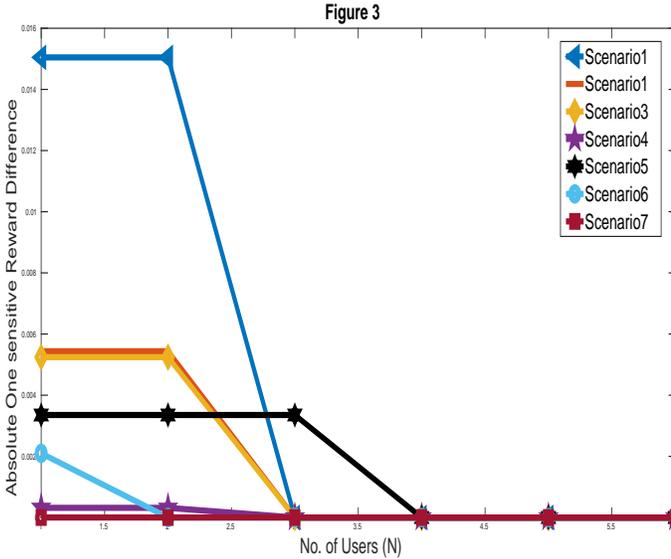}

\caption{Plot of the absolute difference between the One sensitive reward of user $1$ when user $1$ employs its NE policy for the game with $N$ users and the One sensitive reward of user $1$ when user $1$ employs its IINE policy against the total number of user as $N$ varies from $1$ to $N_{max}$.} 
\label{Figure3}
\end{figure}

In Figure \ref{Figure3}, we plot the absolute difference of One-sensitive reawrds of user $1$ computed at the Nash equilibrium policies and the IINE policy of user $1$ versus the number of users(N). Recall that the One-sensitive reward when user $1$ employs policy $z_1$ is \begin{equation}
l_{1}^{1}(z_{1})=\sum_{x_{i},a_{i}}h_{i}p_{i}\cdot z_{i}(x_{i},a_{i}).\label{eq:1th_sensitive_function}
\end{equation}
This is simply the time average SNR of user $1$ at policy $z_1$ and from \ref{thm:Existence_IINE}, the IINE maximizes the time average SNR. In Figure \ref{Figure3}, we plot $|l_1(z_1(N))-l_1(z^*)|$ versus the number of users $N$. 
As observed in Figure \ref{Figure3}, the NE policies maximize the One-sensitive reward of user $1$, once the number of users crosses $N^*$. Indeed, as once the number of users cross $N^*$ the NE policies are infinitely invariant and hence 
maximize the time average SNR. Also in scenario $7$, irrespective of number of users, these policies always are One-sensitive optimal, as here $N^*=1$.

\section{Conclusions}

\vspace{-2mm}

In this paper, we analyzed the scenario where multiple transmitters
can send atmost one packet to a single receiver simultaneously 
over the multiple access channel. We 
model this problem as a Constrained Markov game with independent state
information. That is, each user only knows his own states and actions but only knows the statistics of the state and actions of the other users. Each user selfishly 
maximized their own rate by choosing a power and queue allocation policy subject 
to power and queue constraints. We showed the existence of Nash equilibrium
in this setup and provided an iterative best response algorithm to 
compute this equilibrium for any number of users. We showed that under 
the assumption of ``finitely symmetric users'', there exists an infinitely invariant 
Nash equilibrium; that is, when the total number of users crosses a 
particular threshold ($N^*$), the Nash equilibrium policies of each user 
remains the same. We then showed that an IINE can be computed by solving a finte sequence of Linear programs. 

\appendices{}

\section{Proof of theorem \ref{Ver_ptnl}\label{sec:Proof-of-Potential_thm}}
\begin{IEEEproof}
We first observe that,

\begin{align*}
\begin{aligned}\end{aligned}
 & T_{i}(z_{i},z_{-i})-T_{i}(\hat{z}_{i},z_{-i})\\
=^{1} & \sum_{x_{-i},a_{-i}}\prod_{l\neq i}z_{l}(x_{l},a_{l})\Bigl(\sum_{x_{i},a_{i}}t_{i}(x_{i},a_{i},x_{-i},a_{-i})\\
\cdot & \Bigl(z(x_{i},a_{i})-\hat{z}_{i}(x_{i},a_{i})\Bigl)\Bigr)\\
=^{2} & \sum_{x_{-i},a_{-i}}\prod_{l\neq i}z_{l}(x_{l},a_{l})\cdot\Biggl[\Bigg[\sum_{(x_{i},a_{i})\neq(\hat{x}_{i},\hat{a}_{i})}\\
 & (t_{i}(x_{-i},a_{-i},x_{i},a_{i})\Bigl(z(x_{i},a_{i})-\hat{z}_{i}(x_{i},a_{i})\Bigl)\Bigg]\\
+ & (t_{i}(x_{-i},a_{-i},\hat{x}_{i},\hat{a_{i}})\Bigl(z(\hat{x}_{i},\hat{a_{i}})-\hat{z}_{i}(\hat{x}_{i},\hat{a_{i}})\Bigl)\Bigg]\\
=^{3} & \sum_{x_{-i},a_{-i}}\prod_{l\neq i}z_{l}(x_{l},a_{l})\cdot\Biggl[\sum_{(x_{i},a_{i})\neq(\hat{x}_{i},\hat{a}_{i})}\Big(t_{i}(x_{-i},a_{-i},x_{i},a_{i})\\
- & (t_{i}(x_{-i},a_{-i},\hat{x}_{i},\hat{a_{i}})\Big)\Bigl(z(x_{i},a_{i})-\hat{z}_{i}(x_{i},a_{i})\Bigl)\Bigg]\\
=^{4} & \sum_{x_{-i},a_{-i}}\prod_{l\neq i}z_{l}(x_{l},a_{l})\cdot\Biggl[\sum_{(x_{i},a_{i})\neq(\hat{x}_{i},\hat{a}_{i})}\Big(t(x_{-i},a_{-i},x_{i},a_{i})\\
- & (t(x_{-i},a_{-i},\hat{x}_{i},\hat{a_{i}})\Big)\Bigl(z(x_{i},a_{i})-\hat{z}_{i}(x_{i},a_{i})\Bigl)\Bigg]\\
=^{5} & T(z_{i},z_{-i})-T(\hat{z}_{i},z_{-i}).
\end{align*}

The equality ($5$) in the above proof follows from reversing the
steps of the above result. Equality $3$ follows from the observation
that $\sum_{x_{i},a_{i}}z_{i}(x_{i},a_{i})=\sum_{x_{i},a_{i}}\hat{z}_{i}(x_{i},a_{i})$. 
\end{IEEEproof}

\section{Proof of theorem \ref{thm:NAS_IINE}\label{sec:Proof_NAS_IINE}}

Before, we provide the proof of theorem (\ref{thm:NAS_IINE}), we
give some definition and notation which are used repeatedly. We use
Asymptotic notation, i.e, given two real valued functions, $f(n)$
and $g(n)$on the set of natural numbers, denote, $f(n)=o(g(n))$,
$f(n)=O(g(n))$ and $f(n)=\Theta(g(n))$ whenever there exist constants
$c_{1}$ and $c_{2}$, both strictly greater than $0,$ and a natural
number $N_{0}$ such that for all $n\geq N_{0}$, $f(n)>c_{1}g(n)$,
$f(n)<c_{2}g(n)$ or $c_{1}g(n)<f(n)<c_{2}g(n)$ respectively. We
also denote 
\begin{equation}
\mu=\inf_{i\geq0}\mathbb{E}(X_{i})\,\text{and\,\ensuremath{\beta}\,=\,\ensuremath{\sup_{i\geq0}\mathbb{E}}(\ensuremath{X_{i}}),}\label{eq:UPPER=000026LOWER_MEAN}
\end{equation}
 where the random variable $X_{i}$ is defined as,
\begin{equation}
X_{i}=h_{i}p_{i}\,\text{{w.p} }\,\sum_{q_{i},c_{i}}z_{i}^{*}(h_{i},p_{i},c_{i},q_{i}),\label{eq:SNR_RandomVariable}
\end{equation}
 where $z_{i}^{*}$ is an IINE policy of user $i$. Note that, from
assumptions of Theorem (\ref{thm:NAS_IINE}), we show in Lemma (\ref{lem:Set_of_all_best_resp}) $\mu>0$
and $\beta<\infty$. The following well known inequality is also used
heavily in our work. We mention it without proof. 
\begin{thm}[Hoeffding's Inequality ]
\label{thm:HOEFFDING_THM} We have for some constant $c>0$, 
\end{thm}
\begin{equation}
P\left(\left|\frac{\sum_{j\geq2}^{N+1}\left(X_{j}-\mathbb{E}(X_{j})\right)}{N}\right|\geq t\right)\leq2\exp(-cNt^{2})\label{eq:HOEFFDING}
\end{equation}

We mention that certain Lemmas required in the proof of theorem (\ref{thm:NAS_IINE})
are provided after the statement of the proof. 

$\smallskip$

\begin{IEEEproof}[Proof of Theorem (\ref{thm:NAS_IINE})]

We first show that if $z_{1}^{*}\in\mathcal{S}_{1}$, then $z_{1}^{*}$
is an IINE policy. We have from Lemma (\ref{lem:FEASIBLE_POLICY}),
that $l_{1}^{1}(z_{1}^{*})>0$. Let $z_{1}$ denote any feasible policy
of user $1$ such that it is also a vertex/endpoint of the polyhedron
$\mathcal{Z}_{i}$ and $z_{1}\notin\mathcal{S}_{1}$. We shall prove
that there exist some positive number $N_{k_{1}}$such that for all
$N\geq N_{k_{1}}$, $T_{1}(z_{1}^{*},z_{-1}^{*})-T_{1}(z_{1},z_{-1}^{*})>0$. 

\begin{align}
 & \left(T_{1}(z_{1}^{*},z_{-1}^{*})-T_{1}(z_{1},z_{-1}^{*})\right)\left(N\right)^{k}\nonumber \\
= & \sum_{x_{1},a_{1}}\mathbb{\Bigg[E}\left[\left(N\right)^{k}\log_{2}\left(1+\frac{h_{1}p_{1}1_{q_{1}>0}}{\sum_{j\geq2}^{N+1}X_{j}+N_{0}}\right)\right]\big(z_{1}^{*}(x_{1},a_{1})\label{eq:Converse_1}\\
- & z_{1}(x_{1},a_{1})\big)\Bigg].\nonumber 
\end{align}

As $z_{1}\notin\mathcal{S}_{1}$, there exist a positive integer $k$
such that $z_{1}\notin\mathcal{S}_{1}^{k}$. Let $k$ represent the
smallest such integer. Then $z_{1}^{*}\in\mathcal{S}_{1}^{m},\,1\leq\,m\,\leq k-1$.
Then using a Taylor series expansion in the previous expression (\ref{eq:Converse_1}),
we have for some constant $c$,

\begin{align}
 & T_{1}(z_{1}^{*},z_{-1}^{*})-T_{1}(z_{1},z_{-1}^{*})\nonumber \\
= & \mathbb{E}\left[\frac{\left(N\right)^{k}}{\left(\sum_{j\geq2}^{N+1}X_{j}+N_{0}\right)^{k}}\right]\Bigg[\sum_{x_{1},a_{1}}\left(-1\right)^{k+1}\left(h_{1}p_{1}1_{q_{1}>0}\right)^{k}\nonumber \\
\cdot & \left(z_{1}^{*}(x_{1},a_{1})-z_{1}(x_{1},a_{1})\right)\Bigg]\,+c\mathbb{E}\left[\frac{\left(N\right)^{k}}{\left(\sum_{j\geq2}^{N+1}X_{j}+N_{0}\right)^{k+1}}\right]\nonumber \\
\cdot & \sum_{x_{1},a_{1}}\left(-1\right)^{k}\left(h_{1}p_{1}1_{q_{1}>0}\right)^{k+1}\left(z_{1}^{*}(x_{1},a_{1})-z_{1}(x_{1},a_{1})\right).\label{eq:Converse_2}
\end{align}

Using that $z_{1}^{*}\notin\mathcal{S}_{1}^{k},\,z_{1}^{*}\in\mathcal{S}_{1}^{m},\,1\leq\,m\,\leq k-1$
and Lemma (\ref{lem:HOEFFDING_BD_APPLICATION}), we have in (\ref{eq:Converse_2}),
for some positive constant $c_{1}>0$, and some constant $c_{2}$

\begin{align*}
T_{1}(z_{1}^{*},z_{-1}^{*})-T_{1}(z_{1},z_{-1}^{*}) & >c_{1}-\frac{c_{2}}{N}.
\end{align*}

Hence there exist a positive number $N_{1}$such that for all $N\geq N_{1}$,
$T_{1}(z_{1}^{*},z_{-1}^{*})-T_{1}(z_{1,j},z_{-1}^{*})>0$. Let $N_{j}$
denote the positive number such that for all $N\geq N_{j}$, $T_{1}(z_{1}^{*},z_{-1}^{*})-T_{1}(z_{j},z_{-1}^{*})>0$,
where $z_{,j}$ represents a vertex/endpoint of the polyhedron $\mathcal{Z}_{1}$
and $z_{,j}\notin\mathcal{S}_{1}$. We claim that $T_{1}(z_{1}^{*},z_{-1}^{*})-T_{1}(z_{1},z_{-1}^{*})\geq0$
for all $z_{1}\in\mathcal{Z}_{1}$ ,for all $N\geq N^{*}=\max_{j}N_{1,j}$.
Indeed, to verify (\ref{eq:Best_Rep_LP}), it suffices to consider
only the vertices of the polyhedron $\mathcal{Z}_{1}$ as the optimization
problem (\ref{eq:Best_Rep_LP}) is a Linear program. Clearly, we have
that for $N\geq N^{*}$, $T_{1}(z_{1}^{*},z_{-1}^{*})-T_{1}(z_{j},z_{-1}^{*})>0$,
where $z_{j}$ represents a vertex/endpoint of the polyhedron $\mathcal{Z}_{i}$
and $z_{j}\notin\mathcal{S}_{1}$. Now consider those points $\hat{z}_{1}$,
which belong to the set $\mathcal{S}_{1}$ and are also vertex of
the polyhedron $\mathcal{Z}_{1}$. Then we have, from (\ref{eq:Infty_Sensitive_set}),
$l_{1}^{k}(z_{1}^{*})=l_{1}^{k}(\hat{z}_{1})$, for all $k.$ Hence
if we define the distribution

\[
\hat{P}\left(X_{1}=h_{1}p_{1}\right)=\sum_{q_{1},c_{1}}\hat{z}_{1}(h_{1},p_{1},q_{1},c_{1}),
\]
 then we see that the SNR random variable $X_{1}$ has the same moments
according to the distributions $P$ and $\hat{P}.$ Hence, by method
of moments, we have $P=\hat{P}.$ It can be shown now that $T_{1}(z_{1}^{*},z_{-1}^{*})=T_{1}(\hat{z}_{1,},z_{-1}^{*})$,
for any $\mathcal{N}$. In particular, this implies that $T_{1}(z_{1}^{*},z_{-1}^{*})=T_{1}(\hat{z}_{1,},z_{-1}^{*})$,
for all $N\geq N^{*}$. Hence, we have that $T_{1}(z_{1}^{*},z_{-1}^{*})\geq T_{1}(z_{1,},z_{-1}^{*})$,
for all for all $z_{1}\in\mathcal{Z}_{1}$ and $N\geq N^{*}$, thus
$z_{1}^{*}$ is a IINE policy. 

We now prove the converse. Let $z_{1}^{*}$ denote an IINE policy
for user $i$, we shall prove that $z_{1}^{*}\in\mathcal{S}_{1}$
using induction. As $z_{1}^{*}$ is an IINE policy, we have from definition
(\ref{def:IINE_defn}) and (\ref{eq:SNR_RandomVariable}) that for
any policy $z_{1}\in\mathcal{Z}_{1}$, for all $N\geq N^{*},$

\begin{align}
\begin{aligned}\end{aligned}
 & \sum_{x_{1},a_{1}}\mathbb{\Bigg[{E}}\left[N\log_{2}\left(1+\frac{h_{1}p_{1}1_{q_{1}>0}}{\sum_{j\geq2}^{N+1}X_{j}+N_{0}}\right)\right]\big(z_{1}^{*}(x_{1},a_{1})\label{eq:Condition_1}\\
- & z_{1}(x_{1},a_{1})\big)\Bigg]\,\geq\,0\nonumber 
\end{align}

At sufficiently large $N$, we have using a Taylor series expansion,
for some constant $c>0,$

\begin{align*}
 & \mathbb{E}\left[N\log_{2}\left(1+\frac{h_{1}p_{1}1_{q_{1}>0}}{\sum_{j\geq2}^{N+1}X_{j}+N_{0}}\right)\right]\\
= & \mathbb{\mathbb{E}}\left[\frac{Nh_{1}p_{1}1_{q_{1}>0}}{\sum_{j\geq2}^{N+1}X_{j}+N_{0}}\right]+c\mathbb{\mathbb{E}}\left[N\cdot O\left(\frac{1}{\left(\sum_{j\geq2}^{N+1}X_{j}+N_{0}\right)^{2}}\right)\right]
\end{align*}

Using Lemma (\ref{lem:HOEFFDING_BD_APPLICATION}), the latter is,

\begin{equation}
=\Theta(1)h_{i}p_{i}1_{q_{i}>0}+\Theta\left(\frac{1}{N}\right).\label{eq:Asymptotic_k=00003D1}
\end{equation}

Applying the latter bound (\ref{eq:Asymptotic_k=00003D1}) to get
an upper bound for (\ref{eq:Condition_1}), we have for some constants
$c_{1},c_{2}$, both strictly greater than $0,$

\[
\sum_{x_{i},a_{i}}c_{1}h_{1}p_{1}1_{q_{1}>0}\left(z_{1}^{*}(x_{1},a_{1})-z_{1}(x_{1},a_{1})\right)+\frac{c_{2}}{N}\geq0.
\]

Letting $N\rightarrow\infty$, we get that $z_{i}^{*}\in\mathcal{S}_{1}^{1}.$
We now assume that the result is true till $k-1$, i.e $z_{1}^{*}\in\mathcal{S}_{1}^{m},\,1\leq m\leq k-1$
we show that $z_{1}^{*}\in\mathcal{S}_{1}^{k}.$ We then have by Taylor
series expansion,

\begin{align}
 & \mathbb{E}\left[N^{k}\log_{2}\left(1+\frac{h_{1}p_{1}1_{q_{1}>0}}{\sum_{j\geq2}^{N+1}X_{j}+N_{0}}\right)\right]\nonumber \\
= & c\mathbb{\mathbb{E}}\left[\sum_{l=1}^{k}-N{}^{k}\left(\frac{-h_{1}p_{1}1_{q_{1}>0}}{\sum_{j\geq2}^{N+1}X_{j}+N_{0}}\right)^{l}\right]\nonumber \\
+ & c\mathbb{\mathbb{E}}\left[(N)^{k}\cdot O\left(\frac{1}{\left(\sum_{j\geq2}^{N+1}X_{j}+N_{0}\right)^{K+1}}\right)\right]\nonumber \\
= & c\mathbb{\mathbb{E}}\left[\sum_{l=1}^{k-1}-N{}^{k}\left(\frac{-h_{1}p_{1}1_{q_{1}>0}}{\sum_{j\geq2}^{N+1}X_{j}+N_{0}}\right)^{l}\right]\nonumber \\
+ & \Theta(1)\left(\left(-1\right)^{k+1}h_{i}p_{i}1_{q_{i}>0}\right)^{k}+\Theta\left(\frac{1}{N}\right),\label{eq:Asymptotic_N=00003Dk}
\end{align}

where the last equality follows from Lemma (\ref{lem:HOEFFDING_BD_APPLICATION}).
As $z_{1}^{*}$ is an IINE policy, we have from definition (\ref{def:IINE_defn})
and (\ref{eq:SNR_RandomVariable}) that for any policy $z_{1}\in\mathcal{Z}_{1}$,
for all $N\geq N^{*},$
\begin{align}
\begin{aligned}\end{aligned}
 & \sum_{x_{1},a_{1}}\mathbb{\Bigg[{E}}\left[\left(N\right)^{k}\log_{2}\left(1+\frac{h_{1}p_{1}1_{q_{1}>0}}{\sum_{j\geq2}^{N+1}X_{j}+N_{0}}\right)\right]\big(z_{1}^{*}(x_{1},a_{1})\label{eq:Condition_k}\\
- & z_{1}(x_{1},a_{1})\big)\Bigg]\,\geq\,0.\nonumber 
\end{align}

Using (\ref{eq:Asymptotic_N=00003Dk}) we get a upper bound for the
latter ($\ref{eq:Condition_k}$),Hence we have for some constants
$c_{1},c_{2}$, both strictly greater than $0,$

\begin{align}
 & \sum_{x_{i},a_{i}}\Bigg[c\mathbb{\mathbb{E}}\left[\sum_{l=1}^{k-1}-N{}^{k}\left(\frac{-h_{1}p_{1}1_{q_{1}>0}}{\sum_{j\geq2}^{N+1}X_{j}+N_{0}}\right)^{l}\right]\big(z_{1}^{*}(x_{1},a_{1})\nonumber \\
 & -z_{1}(x_{1},a_{1})\big)\Bigg]\nonumber \\
+ & \sum_{x_{i},a_{i}}c_{1}\left(\left(-1\right)^{k+1}h_{1}p_{1}1_{q_{1}>0}\right)^{k}\left(z_{1}^{*}(x_{1},a_{1})-z_{1}(x_{1},a_{1})\right)\label{eq:Condition_k_1}\\
 & +\frac{c_{2}}{N}\geq0.\nonumber 
\end{align}

As $z_{i}^{*}\in\mathcal{S}_{1}^{m},\,1\leq m\leq k-1,$ then from
(\ref{eq:Sensitive_set_prop}), the first term in the above expression
(\ref{eq:Condition_k_1}) is $0$, hence (\ref{eq:Condition_k_1})
is, 

\begin{align}
= & \sum_{x_{i},a_{i}}c_{1}\left(\left(-1\right)^{k+1}h_{1}p_{1}1_{q_{1}>0}\right)^{k}\left(z_{1}^{*}(x_{1},a_{1})-z_{1}(x_{1},a_{1})\right)\label{eq:Condition_k_2}\\
 & +\frac{c_{2}}{N}\geq0.\nonumber 
\end{align}

Letting $N\rightarrow\infty$, we get that $z_{1}^{*}\in\mathcal{S}_{1}^{k}.$
Hence by induction, $z_{1}^{*}\in\mathcal{S}_{1}$.
\end{IEEEproof}
\begin{lemma}
\label{lem:HOEFFDING_BD_APPLICATION}We have, for each natural number
$k$,

\begin{equation}
\mathbb{\mathbb{E}}\left[\frac{1}{\left(\sum_{j\geq2}^{N+1}X_{j}+N_{0}\right)^{k}}\right]=\Theta\left(\frac{1}{N^{k}}\right).\label{eq:Bound_2}
\end{equation}

\begin{equation}
\mathbb{\mathbb{E}}\left[\frac{N^{k}}{\left(\sum_{j\geq2}^{N+1}X_{j}+N_{0}\right)^{k}}\right]=\Theta(1),\label{eq: Bound_1}
\end{equation}
\end{lemma}
\begin{IEEEproof}
We prove statement (\ref{eq:Bound_2}). Using $X_{j}\leq h_{j}^{k}p_{J}^{l}$,
we have that ,

\begin{equation}
\mathbb{\mathbb{E}}\left[\frac{1}{\left(\sum_{j\geq2}^{N+1}X_{j}+N_{0}\right)^{k}}\right]=o\left(\frac{1}{N^{k}}\right).\label{eq:BOUND2_1}
\end{equation}

To prove the other way, we have 

Proof of theorem \ref{thm:NAS_IINE}\label{sec:Proof_NAS_IINE-1}
\begin{align*}
 & \mathbb{\mathbb{E}}\left[\frac{1}{\left(\sum_{j\geq2}^{N+1}X_{j}+N_{0}\right)^{k}}\right]\\
= & \mathbb{\mathbb{E}}\left[\frac{1_{\left\{ \sum_{j\geq2}^{N+1}X_{j}\geq\frac{1}{2}N\mu\right\} }+1_{\sum_{j\geq2}^{N+1}X_{j}<\frac{1}{2}N\mu}}{\left(\sum_{j\geq2}^{N+1}X_{j}+N_{0}\right)^{k}}\right]\\
\leq & \frac{2^{k}}{\left(N\mu\right)^{k}}+\frac{1}{N_{o}^{k}}P\left(\frac{\sum_{j\geq2}^{N+1}X_{j}}{N}<u/2\right)\\
\leq & \frac{2^{k}}{\left(N\mu\right)^{k}}+\frac{1}{N_{o}^{k}}\exp\left(-c_{1}N\mu^{2}\right)
\end{align*}

where the last inequality follows from Hoeffding's inequality (\ref{eq:HOEFFDING})
. Hence, 

\begin{equation}
\mathbb{\mathbb{E}}\left[\frac{1}{\left(\sum_{j\geq2}^{N+1}X_{j}+N_{0}\right)^{k}}\right]=O\left(\frac{1}{N^{k}}\right),\label{eq:BOUND2.2}
\end{equation}

and (\ref{eq:Bound_2}) follows from (\ref{eq:BOUND2_1}) and (\ref{eq:BOUND2.2}).
We now prove statement (\ref{eq: Bound_1}). Using the proof technique
as carried out for (\ref{eq:Bound_2}), we have, 
\begin{align*}
\frac{1}{\left(h_{j}^{k}p_{J}^{l}\right)^{k}}< & \mathbb{\mathbb{E}}\left[\frac{N^{k}}{\left(\sum_{j\geq2}^{N+1}X_{j}+N_{0}\right)^{k}}\right]<\\
 & \left(\frac{2}{\mu}\right)^{k}+\left(\frac{N}{N_{0}}\right)^{k}\exp\left(-c_{1}N\mu^{2}\right).
\end{align*}

Hence we have (\ref{eq: Bound_1}). 
\end{IEEEproof}
In the next lemma we show that under the assumption that the channel
provides a positive reward for transmission $\left(\pi(h_{i}=0)<1\right)$
and that there is always data to transmit $F_{i}(0)>1$, there always
exists some policy $z_{i}$ such that $l_{i}(z_{i})>0.$ 
\begin{lemma}[Feasible Policy]
\label{lem:FEASIBLE_POLICY} Assume $F_{i}(0)>1$ and $\pi(h_{i}=0)<1$.
Consider the stationary policy $u_{i}$ for user $i$,

\begin{align*}
u_{i}\left(p_{i},c_{i}/h_{i},q_{i}\right) & =\begin{cases}
1 & c_{i}=0,p_{i}=p_{i}^{1},q_{i}\neq0,\forall h_{i}\in\mathcal{H}_{i},\\
1-s & c_{i}=0,p_{i}=0,q_{i}=0,\forall h_{i}\in\mathcal{H}_{i},\\
s & c_{i}=1,p_{i}=0,q_{i}=0,\forall h_{i}\in\mathcal{H}_{i},\\
0 & \text{otherwise},
\end{cases}
\end{align*}

where $p_{i}^{1}=\inf\left\{ p_{i}|p_{i}>0\right\} $. Given any value
of Queue constraint $\overline{Q}_{i}$ and Power constraint $\overline{P}_{i}$,
there exist some value of $s$, under which both the constraints are
satisfied. Also, under the same value of $s$, we have $l_{i}^{1}(z_{i})>0.$
\end{lemma}
\begin{IEEEproof}
Let $z_{i}$ represent the occupation measure corresponding to the
stationary policy $u_{i}$. Let $z_{i}(h_{i})=\sum_{q_{i},a_{i}}z_{i}(h_{i},q_{i},a_{i})$,
$z_{i}(q_{i})=\sum_{h_{i},a_{i}}z_{i}(h_{i},q_{i},a_{i})$ , $z_{i}(p_{i})=\sum_{x_{i},a_{i}}z_{i}(p_{i},c_{i},x_{i})$
and $z_{i}(h_{i},q_{i})=\sum_{a_{i}}z_{i}(h_{i},q_{i},a_{i})$. It
can be easily verified under the policy $u_{i}$, the fading process
$h_{i}[n]$ and the queue process $q_{i}[n]$ are independent. Also
the queue process $q_{i}[n]$ is ergodic with a single communicating
class consisting of the whole set $\mathcal{Q}.$ We denote $\pi(h_{i})$,$\pi(q_{i})$
and $\pi(h_{i},q_{i})$ as the stationary probability of being in
channel state $h_{i}$, stationary probability of queue state $q_{i}$
and the joint stationary probability of state $\left(q_{i},h_{i}\right).$
Then $\pi(h_{i},q_{i})=\pi(h_{i})\pi(q_{i})$, $z_{i}(h_{i})=\pi(h_{i})$,
$z_{i}(q_{i})=\pi(q_{i})$ and $z_{i}(h_{i},q_{i})=\pi(h_{i},q_{i}).$
The transition probability of the queue process under the policy $u_{i}$
is given by, 

\[
P\left(q_{2}/q_{1}\right)=\begin{cases}
sF_{i}(0)+1-s & q_{2}=0,q_{1}=0,\\
sF_{i}(j) & q_{2}=0,q_{1}=j,\,1\leq j\leq Q-1,\\
\sum_{j=Q}^{\infty}sF_{i}(j) & q_{2}=0,q_{1}=Q,\\
1 & q_{2}=j-1,q_{i}=j,1\leq j\leq Q,\\
0 & \text{otherwise.}
\end{cases}
\]

Using the steady state equations, $\pi=\pi P$ for the queue process,
we can show that, 

\begin{align*}
\pi(k) & =s\pi(0)\left(1-\sum_{j=0}^{k-1}F_{i}(j)\right),\\
\pi(0) & =\frac{1}{1+sc}\,,
\end{align*}

where $c=\sum_{k=1}^{Q}\left(1-\sum_{j=0}^{k-1}F_{i}(j)\right)\geq0.$
Note that $c=0$ if and only if $F_{i}(0)=1$ and hence $\pi(0)=1$
if and only if $F_{i}(0)=1$. The average queue length under the policy
$u_{i}$ is,

\begin{align}
Q_{i}(z_{i}) & =\sum_{x_{i},a_{i}}q_{i}z_{i}\left(x_{i},a_{i}\right)\nonumber \\
= & \sum_{q_{i}\neq0}q_{i}\pi(q_{i})\nonumber \\
= & \frac{s}{1+sc}\sum_{k=1}^{Q}\left(\left(1-\sum_{j=0}^{k-1}F_{i}(j)\right)\right)\nonumber \\
\leq & \frac{sQ}{1+sc}\nonumber \\
\leq & sQ.\label{eq:AVG_QUEUE_LENGTH}
\end{align}

The average power expenditure under policy $u_{i}$ is,

\begin{align*}
P_{i}(z_{i}) & =\sum_{x_{i},a_{i}}p_{i}z_{i}(x_{i},a_{i})\\
= & \sum_{q_{i}\neq0}p_{i}^{1}\pi(q_{i})\\
= & \frac{sp_{i}^{1}}{1+sc}\sum_{k=1}^{Q}\left(\left(1-\sum_{j=0}^{k-1}F_{i}(j)\right)\right)\\
\leq & \frac{sQp_{i}^{1}}{1+sc}\\
\leq & sQp_{i}^{1}.
\end{align*}

Now by choosing $0<s<\min\left\{ \frac{\overline{Q}_{i}}{Q},\frac{\overline{P_{i}}}{Qp_{i}^{1}},1\right\} $,
we can ensure that the average queue and power constraint will get
satisfied. We now compute $l_{i}^{1}(z_{i})$ for the policy $u_{i}.$
Define $h_{i}^{1}=\inf\left\{ h_{i}|h_{i}>0\right\} $

\begin{align}
l_{i}^{1}(z_{i}) & =\sum_{x_{i},a_{i}}h_{i}p_{i}z_{i}(x_{i},a_{i})\nonumber \\
\geq & h_{i}^{1}p_{i}^{1}\left(\sum_{q_{i}\neq0}\sum_{h_{i}\neq0}\sum_{p_{i}\neq0}\sum_{c_{i}=0}^{1}z_{i}(h_{i},q_{i},p_{i},c_{i})\right)\nonumber \\
= & h_{i}^{1}p_{i}^{1}\left(\sum_{q_{i}\neq0}\sum_{h_{i}\neq0}\sum_{p_{i}\neq0}{}_{}z_{i}(h_{i},q_{i},p_{i})\right).\label{eq:One_sensitve_rwrd_calculation_1}
\end{align}

As user $i$, always transmit when his queue is not empty, we have
$z_{i}(h_{i},q_{i},p_{i}=0)=0$, for all $q_{i}\neq0$. Thus, we have
in (\ref{eq:One_sensitve_rwrd_calculation_1}) ,

\begin{align*}
l_{i}^{1}(z_{i}) & \geq h_{i}^{1}p_{i}^{1}\left(\sum_{q_{i}\neq0}\sum_{h_{i}\neq0}\sum_{p_{i}}{}_{}z_{i}(h_{i},q_{i},p_{i})\right)\\
= & h_{i}^{1}p_{i}^{1}\left(\sum_{q_{i}\neq0}\sum_{h_{i}\neq0}z_{i}(h_{i},q_{i})\right)\\
= & h_{i}^{1}p_{i}^{1}\left(1-\pi(q_{i}=0)\right)\left(1-\pi(h_{i}=0)\right).
\end{align*}

Thus if we ensure $\pi(q_{i}=0)<1$(or equivalently $F_{i}(0)>1$)
and $\pi(h_{i}=0)<1$, then $l_{i}^{1}(z_{i})>0$. 
\end{IEEEproof}
\begin{lemma}
\label{lem:Set_of_all_best_resp}
Let $z_{i}^{*}$ denote an IINE policy of user $i$ and let the random
variable $X_{i}$ be,
\begin{equation}
X_{i}=h_{i}p_{i}\,\text{{w.p} }\,\sum_{q_{i},c_{i}}z_{i}^{*}(h_{i},p_{i},c_{i},q_{i}),\label{eq:SNR_RandomVariable-1}
\end{equation}
 Then
\begin{equation}
\mu=\inf_{i\geq0}\mathbb{E}(X_{i})>0\,\text{and\,\ensuremath{\beta}\,=\,\ensuremath{\sup_{i\geq0}\mathbb{E}}(\ensuremath{X_{i}})<\ensuremath{\infty}.}\label{eq:UPPER=000026LOWER_MEAN-1}
\end{equation}
\end{lemma}
\begin{IEEEproof}
As the IINE policy is a NE policy, we shall prove that $l_{i}^{1}(z_{i})=\sum_{x_{i},a_{i}}h_{i}p_{i}z_{i}(x_{i},a_{i})>0$
for any best response policy $z_{i}$. To do so, we first define the
set of all best response of user $i$ as,

\begin{equation}
{\mathcal{B}}_{i}=\bigcup_{{\mathcal{N}}\subseteq{\mathbb{Z}}^{+}}\bigcup_{z_{-i}\in{\mathcal{Z}}_{-i}^{\mathcal{N}}}{\mathcal{B}}_{i}(z_{-i}),\label{eq:Set_ALL_best_responses}
\end{equation}

where ${\mathcal{Z}}_{-i}^{\mathcal{N}}$ denotes the the set of all
policies of users other than $i$, when the set ${\mathcal{N}}$ containing
the number $i$ in the game $\Gamma_{\mathcal{N}}$. Let ${\mathcal{C}}({\mathcal{Z}}_{i})$
denote the class of all the subsets of the set ${\mathcal{Z}}_{i}$,
which are obtained as convex closure of finitely many vertices of
the polyhedron ${\mathcal{Z}}_{i}$. As the vertices of the polyhedron
${\mathcal{Z}}_{i}$ is finite, we have the class ${\mathcal{C}}({\mathcal{Z}}_{i})$
itself as finite. We note that the set ${\mathcal{B}}_{i}(z_{-i})$
contains all the best response policies of user $i$, when users other
than user $i$ play multi-policy $z_{-i}$. As this set contains all
the set of solutions of a linear program, we have that this set is
the convex closure of finitely many points, each point being a vertex
of the set ${\mathcal{Z}}_{i}$ of feasible occupation measures of
user $i$. This implies that the best response set ${\mathcal{B}}_{i}(z_{-i})$
belongs to the class ${\mathcal{C}}({\mathcal{Z}}_{i})$ for each
multi-policy $z_{-i}$ of users other than user $i$. As the class
${\mathcal{C}}({\mathcal{Z}}_{i})$ is finite, we have that the set
${\mathcal{B}}_{i}$ is a finite union of compact sets. Hence ${\mathcal{B}}_{i}$
is compact. As the IINE belongs to the set ${\mathcal{B}}_{i}$, it
suffices to show that $\inf_{z_{i}\in{\mathcal{B}}_{i}}l_{i}(z_{i})>0$.
As the set ${\mathcal{B}_{i}}$is compact, we simply show that $l_{i}(z_{i}^{*})>0$
for any policy $z_{i}^{*}\in{\mathcal{B}}_{i}.$ Let $z_{i}^{*}\in{\mathcal{B}}_{i}$
be any best response policy of user $i$ for some game $\Gamma_{\mathcal{N}}$,
when the other users employ policy $z_{-i}$. Let $z_{i}$ denote
any arbitrary policy of user $i$. Then we have that $T_{i}(z_{i},z_{-i})>0$
if and only if $l_{i}^{1}(z_{i})>0$. By lemma (\ref{lem:FEASIBLE_POLICY}),
we have a policy $z_{i}^{1}$ such that $l_{i}^{1}(z_{i}^{1})>0$,
hence we have,$T_{i}(z_{i}^{*},z_{-i})\geq T_{i}(z_{i}^{1},z_{-i})>0$.
Thus $l_{i}(z_{i}^{*})>0$ for the policy $z_{i}^{*}\in{\mathcal{B}}_{i}.$
This shows that $\mu>0$. 
\end{IEEEproof}

\section{Proof of Theorem \ref{thm:Existence_IINE} and Theorem \ref{thm:-Interchangeability_IINE}\label{sec:Appendix_C}}
\begin{IEEEproof}[Proof of Theorem (\ref{thm:Existence_IINE})]
 We note that by assumptions in Theorem (\ref{thm:NAS_IINE}), we
have that $\mathcal{Z}_{i}$ is non-empty, hence there exist a point
$z_{i}\in\mathcal{S}_{i}^{1}.$ Hence by, statement (\ref{eq:Kth_sensitive_set}),
the sets $\mathcal{S}_{i}^{k}$ are nonempty. By property (\ref{eq:Sensitive_set_prop}),
the set $\mathcal{S}_{i}$ is non-empty. Thus by theorem (\ref{thm:NAS_IINE}),
there exist an IINE. We now show that $\mathcal{S}_{i}=\mathcal{S}_{i}^{k*}$,
where $M=\#\left(\left\{ h_{i}p_{i}|h_{i}\in\mathcal{H}_{i}\,,\,p_{i}\in\mathcal{P}_{i}\right\} \right)$.
Let $z_{i}$ and $\hat{z}_{i}$ represent two distinct policies belonging
to the set $\mathcal{S}_{i}^{k}$ and $\mathcal{S}_{i}$ respectively.
Hence, we have that both policies belong to $\mathcal{S}_{i}^{k}$.
We shall now show that $z_{i}\in\mathcal{S}_{i}$. Let $X_{i}$ denote
the SNR random variable which take values in the set $\left\{ h_{i}p_{i}|h_{i}\in\mathcal{H}_{i}\,,\,p_{i}\in\mathcal{P}_{i}\right\} $.
We order the set as $\left\{ x_{1},x_{2},\cdots,x_{M}\right\} $,
with $x_{i}\leq x_{i+1}$, $x_{1}=0$ and $x_{M}=h_{i}^{k}p_{i}^{l}$.
Define two probability distributions, $P$ and $\hat{P}$ such that,

\[
\hat{P}\left(X_{1}=h_{1}p_{1}\right)=\sum_{q_{1},c_{1}}\hat{z}_{1}(h_{1},p_{1},q_{1},c_{1})\,\text{and}
\]

\begin{equation}
P\left(X_{1}=h_{1}p_{1}\right)=\sum_{q_{1},c_{1}}z_{1}(h_{1},p_{1},q_{1},c_{1}).\label{eq:SNR_distbn_defn}
\end{equation}

Let $m_{k}$ and $\hat{m}_{k}$ represent the $k$th moments of the
random variable $X_{i}$ with respect to the two distributions $P$
and $\hat{P}$. As $z_{i}$ and $\hat{z}_{i}$ represent two distinct
policies belonging to the set $\mathcal{S}_{i}^{k}$, we have $m_{k}=\hat{m}_{k},\,\forall\,1\leq k\leq M.$
If we define a matrix $V$ of size $(M-1\times M-1)$ with
entries $V_{k,l}=\left(x_{k}\right)^{l},\,2\leq l\leq M-1,\,2\leq k\leq M-1$,
then we have $V\left(\hat{y}-y\right)=0$, where the $M-1$ vectors
are defined as $\hat{z}=\left(\hat{P}(x_{2}),\hat{P}(x_{3}),\cdots,\hat{P}(x_{M})\right)$
and $z=\left(P(x_{2}),P(x_{3}),\cdots,P(x_{M})\right)$ respectively.
However, as $V$ is an invertible matrix, we have $\hat{y}=y$ and
hence $\hat{P}=P.$ Now as the distributions are the same, this implies
that the moments $m_{k}$ and $\hat{m}_{k}$ are the same for all
$k$. As $l_{i}^{k}(z_{i})=\left(-1\right)^{k+1}m_{k}$ and $l_{i}^{k}(\hat{z}_{i})=\left(-1\right)^{k+1}\hat{m}_{k}$,
we have that $z_{i}\in\mathcal{S}_{i}$ and hence $\mathcal{S}_{i}^{k}\in\mathcal{S}_{i}$.
\end{IEEEproof}
\begin{IEEEproof}[Proof of Theorem (\ref{thm:-Interchangeability_IINE})]
 Let $z_{i}$ and $\hat{z}_{i}$ represent two IINE policies for
each user $i$. Define two probability distributions $P$ and $\hat{P}$
as in (\ref{eq:SNR_distbn_defn}). We can now show by a computation
that for each set $\mathcal{N}$

\[
T_{i}(z_{i},z_{-i})=\mathbb{E}\left[\log_{2}\left(1+\frac{X_{i}}{N_{0}+\sum_{j\neq i}X_{j}}\right)\right],
\]

where the $X_{i}$ are SNR random variables which takes values in
the set $\left\{ h_{i}p_{i}|h_{i}\in\mathcal{H}_{i}\,,\,p_{i}\in\mathcal{P}_{i}\right\} .$
As $z_{i}$ and $\hat{z}_{i}$ are both IINE policies for user $i$,
by using a similar argument as done in the proof of Theorem (\ref{thm:Existence_IINE}),
we have that , and hence $T_{i}(z_{i},z_{-i})=T_{i}(\hat{z}_{i},\hat{z}_{-i})$.
Thus the IINE policies are interchangeable.
\end{IEEEproof}

\vspace{-3mm}
\bibliography{GLOBECOMM_REFRENCES}
\bibliographystyle{ieeetr}

\vspace{-1mm}

\end{document}